\documentclass[twocolumn,showpacs,superscriptaddress,prl]{revtex4}

\usepackage[dvips]{color}
\usepackage{epsfig}

\begin{document}  

\title{Investigation of Proton-Proton Short-Range Correlations via the $^{12}$C(e,e$^{\prime}$pp) Reaction} 

\author{R. Shneor}
\affiliation{Tel Aviv  University, Tel Aviv 69978, Israel} 
\author{P.~Monaghan}
\affiliation{Massachusetts Institute of Technology, Cambridge, Massachusetts 02139, USA}
\author{R.~Subedi}
\affiliation{Kent State University, Kent, Ohio 44242, USA}
\author{B.~D.~Anderson}
\affiliation{Kent State University, Kent, Ohio 44242, USA}
\author{K. Aniol}
\affiliation{California State University Los Angeles, Los Angeles, California 90032, USA}
\author{J.~Annand}
\affiliation{University of Glasgow, Glasgow G12 8QQ, Scotland, UK}
\author{J.~Arrington}
\affiliation{Argonne National Laboratory, Argonne, Illinois, 60439, USA}
\author{H.~Benaoum}
\affiliation{Syracuse University, Syracuse, New York 13244, USA}
\author{F.~Benmokhtar}
\affiliation{University of Maryland, College Park. Maryland 20742, USA}
\author{P.~Bertin}
\affiliation{Laboratoire de Physique Corpusculaire, F-63177 Aubi\`{e}re, France}
\author{W.~Bertozzi}
\affiliation{Massachusetts Institute of Technology, Cambridge, Massachusetts 02139, USA}
\author{W.~Boeglin}
\affiliation{Florida International University, Miami, Florida 33199, USA}
\author{J.~P.~Chen}
\affiliation{Thomas Jefferson National Accelerator Facility, Newport News, Virginia 23606, USA}
\author{Seonho~Choi}
\affiliation{Seoul National University, Seoul 151-747, Korea}
\author{E.~Chudakov}
\affiliation{Thomas Jefferson National Accelerator Facility, Newport News, Virginia 23606, USA}
\author{E.~Cisbani}
\affiliation{INFN, Sezione Sanit\'{a} and Istituto Superiore di Sanit\'{a}, Laboratorio di Fisica, I-00161 Rome, Italy}
\author{B.~Craver}
\affiliation{University of Virginia, Charlottesville, Virginia 22904, USA}
\author{C.~W.~de~Jager}
\affiliation{Thomas Jefferson National Accelerator Facility, Newport News, Virginia 23606, USA}
\author{R.~Feuerbach}
\affiliation{Thomas Jefferson National Accelerator Facility, Newport News, Virginia 23606, USA}
\author{S.~Frullani}
\affiliation{INFN, Sezione Sanit\'{a} and Istituto Superiore di Sanit\'{a}, Laboratorio di Fisica, I-00161 Rome, Italy}
\author{F.~Garibaldi}
\affiliation{INFN, Sezione Sanit\'{a} and Istituto Superiore di Sanit\'{a}, Laboratorio di Fisica, I-00161 Rome, Italy}
\author{O.~Gayou}
\affiliation{Massachusetts Institute of Technology, Cambridge, Massachusetts 02139, USA}
\author{S.~Gilad}
\affiliation{Massachusetts Institute of Technology, Cambridge, Massachusetts 02139, USA}
\author{R.~Gilman}
\affiliation{Rutgers, The State University of New Jersey, Piscataway, New Jersey 08855, USA}
\affiliation{Thomas Jefferson National Accelerator Facility, Newport News, Virginia 23606, USA}
\author{O.~Glamazdin}
\affiliation{Kharkov Institute of Physics and Technology, Kharkov 310108, Ukraine}
\author{J.~Gomez}
\affiliation{Thomas Jefferson National Accelerator Facility, Newport News, Virginia 23606, USA}
\author{O.~Hansen}
\affiliation{Thomas Jefferson National Accelerator Facility, Newport News, Virginia 23606, USA}
\author{D.~W.~Higinbotham}
\affiliation{Thomas Jefferson National Accelerator Facility, Newport News, Virginia 23606, USA}
\author{T.~Holmstrom}
\affiliation{College of William and Mary, Williamsburg, Virginia 23187, USA}               
\author{H.~Ibrahim}
\affiliation{Old Dominion University, Norfolk, Virginia 23508, USA}
\author{R.~Igarashi}
\affiliation{University of Saskatchewan, Saskatoon, Saskatchewan, Canada S7N 5E2}
\author{E.~Jans}
\affiliation{Nationaal Instituut voor Kernfysica en Hoge-Energiefysica, Amsterdam, The Netherlands}
\author{X.~Jiang}
\affiliation{Rutgers, The State University of New Jersey, Piscataway, New Jersey 08855, USA}
\author{Y.~Jiang}
\affiliation{University of Science and Technoogy of China, Hefei, Anhui, China}
\author{L.~Kaufman}
\affiliation{University of Massachusetts Amherst, Amherst, Massachusetts 01003, USA}
\author{A.~Kelleher}
\affiliation{College of William and Mary, Williamsburg, Virginia 23187, USA}               
\author{A.~Kolarkar}
\affiliation{University of Kentucky, Lexington, Kentucky 40506, USA}
\author{E.~Kuchina}
\affiliation{Rutgers, The State University of New Jersey, Piscataway, New Jersey 08855, USA}
\author{G.~Kumbartzki} 
\affiliation{Rutgers, The State University of New Jersey, Piscataway, New Jersey 08855, USA}
\author{J.~J.~LeRose}
\affiliation{Thomas Jefferson National Accelerator Facility, Newport News, Virginia 23606, USA}
\author{R.~Lindgren}
\affiliation{University of Virginia, Charlottesville, Virginia 22904, USA}
\author{N.~Liyanage}
\affiliation{University of Virginia, Charlottesville, Virginia 22904, USA}
\author{D.~J.~Margaziotis}
\affiliation{California State University Los Angeles, Los Angeles, California 90032, USA}
\author{P.~Markowitz}
\affiliation{Florida International University, Miami, Florida 33199, USA}
\author{S.~Marrone}
\affiliation{INFN, Sezione Sanit\'{a} and Istituto Superiore di Sanit\'{a}, Laboratorio di Fisica, I-00161 Rome, Italy}
\author{M.~Mazouz}
\affiliation{Laboratoire de Physique Subatomique et de Cosmologie, 38026 Grenoble, France}
\author{R.~Michaels}
\affiliation{Thomas Jefferson National Accelerator Facility, Newport News, Virginia 23606, USA}
\author{B.~Moffit}
\affiliation{College of William and Mary, Williamsburg, Virginia 23187, USA}               
\author{S.~Nanda}
\affiliation{Thomas Jefferson National Accelerator Facility, Newport News, Virginia 23606, USA}
\author{C.~F.~Perdrisat}
\affiliation{College of William and Mary, Williamsburg, Virginia 23187, USA}               
\author{E.~Piasetzky}
\affiliation{Tel Aviv  University, Tel Aviv 69978, Israel} 
\author{M.~Potokar} 
\affiliation{Institute ``Jo\v{z}ef Stefan'', 1000 Ljubljana, Slovenia}
\author{V.~Punjabi}
\affiliation{Norfolk State University, Norfolk, Virginia 23504, USA}
\author{Y.~Qiang}
\affiliation{Massachusetts Institute of Technology, Cambridge, Massachusetts 02139, USA}
\author{J.~Reinhold}
\affiliation{Florida International University, Miami, Florida 33199, USA}
\author{B.~Reitz}
\affiliation{Thomas Jefferson National Accelerator Facility, Newport News, Virginia 23606, USA}
\author{G.~Ron}
\affiliation{Tel Aviv  University, Tel Aviv 69978, Israel} 
\author{G.~Rosner}
\affiliation{University of Glasgow, Glasgow G12 8QQ, Scotland, UK}
\author{A.~Saha} 
\affiliation{Thomas Jefferson National Accelerator Facility, Newport News, Virginia 23606, USA}
\author{B.~Sawatzky}
\affiliation{University of Virginia, Charlottesville, Virginia 22904, USA}
\affiliation{Temple University, Philadelphia, Pennsylvania 19122, USA}
\author{A.~Shahinyan}
\affiliation{Yerevan Physics Institute, Yerevan 375036, Armenia}
\author{S.~\v{S}irca}
\affiliation{Institute ``Jo\v{z}ef Stefan'', 1000 Ljubljana, Slovenia}
\affiliation{Dept. of Physics, University of Ljubljana, 1000 Ljubljana, Slovenia}
\author{K.~Slifer}
\affiliation{University of Virginia, Charlottesville, Virginia 22904, USA}
\affiliation{Temple University, Philadelphia, Pennsylvania 19122, USA}
\author{P.~Solvignon}
\affiliation{Temple University, Philadelphia, Pennsylvania 19122, USA}
\author{V.~Sulkosky}
\affiliation{College of William and Mary, Williamsburg, Virginia 23187, USA}               
\author{N.~Thompson}
\affiliation{University of Glasgow, Glasgow G12 8QQ, Scotland, UK}
\author{P.~E.~Ulmer}
\affiliation{Old Dominion University, Norfolk, Virginia 23508, USA}
\author{G.~M.~Urciuoli}
\affiliation{INFN, Sezione Sanit\'{a} and Istituto Superiore di Sanit\'{a}, Laboratorio di Fisica, I-00161 Rome, Italy}
\author{E.~Voutier}
\affiliation{Laboratoire de Physique Subatomique et de Cosmologie, 38026 Grenoble, France}
\author{K.~Wang}
\affiliation{University of Virginia, Charlottesville, Virginia 22904, USA}
\author{J.~W.~Watson}
\affiliation{Kent State University, Kent, Ohio 44242, USA}
\author{L.B.~Weinstein} 
\affiliation{Old Dominion University, Norfolk, Virginia 23508, USA}
\author{B.~Wojtsekhowski}
\affiliation{Thomas Jefferson National Accelerator Facility, Newport News, Virginia 23606, USA}
\author{S.~Wood} 
\affiliation{Thomas Jefferson National Accelerator Facility, Newport News, Virginia 23606, USA}
\author{H.~Yao}
\affiliation{Temple University, Philadelphia, Pennsylvania 19122, USA}
\author{X.~Zheng}
\affiliation{Argonne National Laboratory, Argonne, Illinois, 60439, USA}
\affiliation{Massachusetts Institute of Technology, Cambridge, Massachusetts 02139, USA}
\author{L.~Zhu}
\affiliation{University of Illinois at Urbana-Champaign, Urbana, Illinois 61801, USA}
\collaboration{The Jefferson Lab Hall A Collaboration}
\noaffiliation

\date{\today}

\begin{abstract}
We investigated simultaneously the $^{12}$C(e,e$^{\prime}$p) and $^{12}$C(e,e$^{\prime}$pp) 
reactions at $Q^{2}$~=~2~(GeV/c)$^{2}$, $x_{B}$ = 1.2, and in an (e,e$^{\prime}$p) missing-momentum
range from 300 to 600~MeV/c.   At these kinematics, with a missing-momentum greater than the Fermi momentum 
of nucleons in a nucleus and far from the delta excitation, short-range nucleon-nucleon correlations are predicted to dominate the reaction.  
For $(9.5\pm2)\%$ of the $^{12}$C(e,e$^{\prime}$p) events, a recoiling partner proton was observed 
back-to-back to the $^{12}$C(e,e$^{\prime}$p) missing momentum vector, an experimental 
signature of correlations.  
\end{abstract}

\maketitle

The short-range component of the nucleon-nucleon force manifests itself via nucleon pairs
inside a nucleus.
Such nucleon pairs have a low center-of-mass momentum and a high relative 
momentum~\cite{Frankfurt:1981mk} where low and high are relative
to the Fermi sea level, $k_{F}$, which for $^{12}$C is 
$\sim220$~MeV/c~\cite{Moniz:1971mt}. 
We refer to such a proton pair as a proton-proton short-range 
correlation (pp-SRC).  
Averaged over all nucleon momenta, the probability for a nucleon in $^{12}$C to be a member of a two-nucleon SRC 
state, proton-proton (pp), proton-neutron (pn), or neutron-neutron (nn), 
has been estimated from the dependence of inclusive (e,e$^{\prime}$) data on the Bjorken scaling
variable, $x_B$, to be
$20\pm 5\%$~\cite{Frankfurt:1993sp,Egiyan:2003vg,Egiyan:2005hs}.   
Measurements at Brookhaven National Laboratory (BNL) of (p,pp) and (p,ppn) at
high momentum transfer ~\cite{Aclander:1999fd,Tang:2002ww,Malki:2000gh}
verified the existence of correlated np pairs; 
subsequent analysis of these data set an upper limit of 3\% for pp-SRCs in $^{12}$C~\cite{Piasetzky:2006ai}.

Even though the probability of pp-SRCs in nuclei is small, they 
are important since they can teach us about the strong interaction at
short distances.  Moreover, as a manifestation of asymmetric dense cold nuclear matter that 
can be studied in the laboratory, they are relevant to the understanding of
neutron stars~\cite{Sargsian:2002wc}.

In this work, we determine the 
fraction of $^{12}$C(e,e$^{\prime}$p) events which are associated with pp-SRC pairs.
This was done by measuring the ratio of the $^{12}$C(e,e$^{\prime}$pp) and the $^{12}$C(e,e$^{\prime}$p) 
cross sections
as a function of the (e,e$^{\prime}$p) missing momentum, $p_{miss}$,
where $\vec p_{miss}$ = $\vec p$ - $\vec q$ as shown in Fig.~1.
By measuring above the Fermi sea of nucleon motion, i.e. greater than 220~MeV/c, and in kinematics where
other reaction mechanisms are suppressed, if the initial struck nucleon is part of a pair, one would expect a single
recoil nucleon to balance the missing momentum vector.
In the impulse approximation, a virtual photon with a large $Q^2$ 
is absorbed by one of the protons in the pair.  This supplies the 
energy required to break the pair and remove the two protons from the nucleus.
Pre-existing pairs are identified by detecting a recoiling proton 
in coincidence with an (e,e$^{\prime}$p) event, where the recoiling proton has a high momentum 
($\vec p_{rec}$) opposite to the direction but of roughly equal magnitude 
to $\vec p_{miss}$ (see Fig.~1).  
  
Historically, the interpretation of triple-coincidence 
data in terms of SRCs has been plagued 
by contributions from meson-exchange currents (MECs), 
isobar configurations (ICs) and 
final-state interactions (FSIs)~\cite{PhysRevLett.74.1712,Blomqvist:1998gq,Groep:2000cy}. 
The kinematics for the measurements described here were chosen to 
minimize these effects.  For example, at high $Q^2$, MEC contributions  
decrease as $1/Q^2$ relative to PWIA contributions and are reduced relative to those due to SRC~\cite{Arnold:1989qr,Laget:1987hb}.  
A large $Q^2$ and $x_B$ also drastically reduces IC contributions~\cite{Frankfurt:1996xx,Sargsian:2001ax}. 
Finally, FSIs are minimized by having a large $\vec p_{miss}$
component antiparallel to the virtual photon direction~\cite{Frankfurt:1996xx}.

This experiment was performed in Hall A of the Thomas Jefferson
National Accelerator Facility (JLab) using an incident electron beam
of 4.627 GeV with a current between  5 and 40~$\mu$A.  The target was a 0.25~mm thick 
graphite sheet rotated 70$^{\circ}$ from perpendicular to the beam line to minimize 
the material through which the recoiling protons passed.
The two Hall A 
high-resolution spectrometers (HRS)~\cite{Alcorn:2004sb} were used to identify the $^{12}$C(e,e$^{\prime}$p) 
reaction.  Scattered electrons were detected in the left HRS (HRS-L) at a
central scattering angle (momentum) of 19.5$^{\circ}$
(3.724 GeV/c). This corresponds to the quasi-free knockout of a single proton with
transferred three-momentum $| \vec{q} |$ = 1.65 GeV/c, transferred energy
$\omega=0.865$ GeV, $Q^2=2$ (GeV/c)$^2$, and
$x_{B}\equiv{Q^{2}\over 2m\omega}=1.2$ where $m$ is the mass of a proton.
Knocked-out 
protons were detected using the right HRS (HRS-R) which was set
at 3 different combinations of central angle and momentum: 40.1$^{\circ}$ \& 
1.45 GeV/c, 35.8$^{\circ}$ \& 1.42 GeV/c, and 32.0$^{\circ}$ \& 1.36 GeV/c.  These
kinematic settings correspond to median missing-momentum values $p_{miss}$ = 0.35,
0.45 and 0.55 GeV/c, respectively, with a range of approximately $\pm 50$ MeV/c each.

A third, large-acceptance spectrometer, BigBite, was used to detect recoiling protons 
in the $^{12}$C(e,e$^{\prime}$pp) events.  
The BigBite spectrometer~\cite{bigbite} consists of a 
large-acceptance, non-focusing  dipole magnet and a detector
package.  For this measurement, the magnet was located at an angle
of 99$^{\circ}$ and 1.1~m from 
the target with a resulting angular acceptance of about 96~msr and 
a nominal momentum acceptance from 0.25 GeV/c to 0.9~GeV/c. 
The detector package was constructed specifically for this experiment. 
It consisted of three planes
of plastic scintillator segmented in the dispersive direction.
The first scintillator plane (the "auxiliary plane"), was placed
at the exit of the dipole,  parallel to the magnetic field boundary,
and consisted of 56 narrow scintillator bars of dimension 350 x 25 x
2.5 mm$^3$.  
The second and third scintillator planes, known collectively as the trigger plane, were mounted together
and were located 1 meter downstream of the first plane. The second and third planes consisted
of 24 scintillator bars each, with dimensions $500\times86\times3$~mm$^3$ and $500\times86\times30$~mm$^3$, 
respectively. 
The scintillator bars in these two layers were offset from one another by half a bar in the dispersive direction, 
improving their position resolution by a factor of two.
Each of the scintillator bars in the auxiliary plane was read out by one photomultiplier tube (PMT), 
while each of the trigger scintillators was read out by two PMTs, one on each end.  
This unshielded system was able to run in Hall~A up to a luminosity of 
$10^{38}$~cm$^{-2}$s$^{-1}$ per nucleon.

\begin{figure}[ht]
\label{triple-fig1}
\centerline{\epsfig{width=\linewidth,file=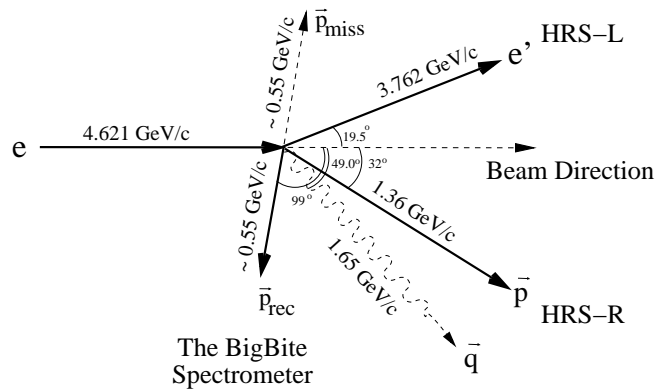}}
\caption{A vector diagram of the layout of the $^{12}$C(e,e$^{\prime}$pp) experiment shown for
the largest $p_{miss}$ kinematics of 0.55~GeV/c.}
\end{figure}

The coincident $^{12}$C(e,e$^{\prime}$p) events  were 
detected in the two HRSs, with a typical trigger rate of 0.2~Hz. 
After spectrometer acceptance cuts,
the time-difference distribution showed a clear electron-proton
coincidence peak with a width of $\sim$ 0.5~ns sigma. 
The measured $^{12}$C(e,e$^{\prime}$p) missing-energy
spectrum for the lowest missing-momentum setting 
($p_{miss} \sim 0.35$ GeV/c) is shown in Fig.~2.  
Missing energy is defined by $E_{miss}\equiv\omega-T_{p}-T_{A-1}$, where
$T_{p}$ is the measured kinetic energy of the knocked-out proton and $T_{A-1}$ is the calculated
kinetic energy of the residual A-1 system. 
The contribution of missing energy due to a single 
proton removal from the $p$-shell in $^{12}$C, leaving the $^{11}$B nucleus 
in its ground state, is seen as a peak at missing energy of about 16 MeV. 
The strength above the $^{11}$B ground state is comprised of $p$-shell removal to highly-excited bound states
and $p$-shell and $s$-shell removal to the continuum. 
The contribution due to $\Delta$-resonance excitation was removed 
by requiring $\vec p_{miss}$ to point in the direction
one would expect from the break-up of a pair, e.g. $< 76^{\circ}$, $< 84^{\circ}$,
and  $< 88^{\circ}$ for the three kinematics, respectively.
This cut removes the
$\Delta$-resonance, since the missing-momentum vector for pion production events 
by conservation of energy and momentum points to
larger angles than direct knock-out events.
The measured missing-energy spectrum 
with and without this angular cut is shown in Fig.~2.

The BigBite spectrometer was positioned to determine if a single high-momentum proton
was balancing the $p_{miss}$ of the (e,e$^{\prime}$p) reaction. 
Such recoiling protons were identified in 
BigBite using the measured energy loss in the scintillator detectors and the consistency 
between the measured time-of-flight (TOF) and the momentum measured by the trajectory in the
magnetic field. The momentum 
resolution of BigBite,  determined from elastic electron-proton
scattering, was ${\Delta p \over p}=4\%$. 
The singles rates with a 30~$\mu$A beam were about 100 kHz per scintillator in the first plane 
and 80 kHz per scintillator in the third plane. 
With these rates, nearly all events had only one track  
with a reconstructed momentum consistent
with the momentum from the TOF.
For the small number of events that had more than one possible reconstructed track,
we selected the track that had the most consistent momentum between the TOF determination and from
ray tracing.  
Primarily due to the gaps between scintillators, the overall proton detection efficiency was 85\%. 

The TOF for protons detected in BigBite was defined from the target to the third 
scintillator plane ($\sim 3$~m) assuming the protons leave the center 
of the target at the same time as the scattered electrons and the knocked-out protons
and was corrected using the reconstructed trajectory path length.
The timing peak shown 
in the insert of Fig.~2 is thus 
due to real triple coincidences and the flat background is due to random 
coincidences between the $^{12}$C(e,e$^{\prime}$p) reaction and protons in BigBite.
The use of a proton identification cut, an angular acceptance cut in
BigBite, and a TOF cut of $\pm 3.5$ ns to select the real coincidences, 
resulted in signal/background ratios of 1:2, 1:1, and 2:1 for 
the median missing-momentum settings of 0.35, 0.45 and 0.55 GeV/c, respectively. 

\begin{figure}[ht]
\centerline{\epsfig{width=\linewidth,file=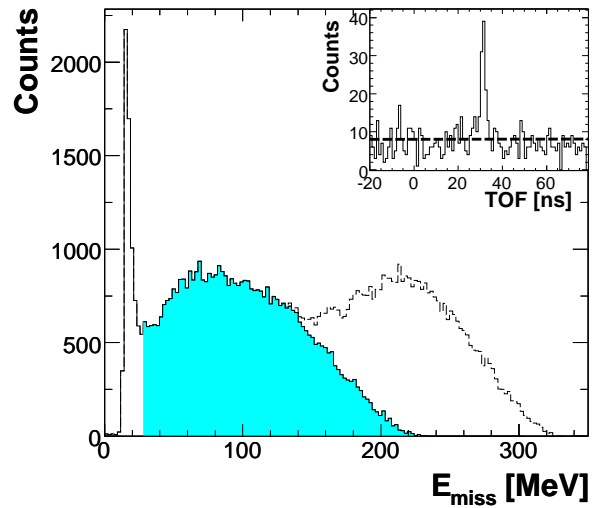}}
\caption{The measured $^{12}$C(e,e$^{\prime}$p) missing-energy spectrum 
for ${p}_{miss} \sim 0.31$~GeV/c. The peak at 16 MeV is due to
removal of $p$-shell protons leaving the $^{11}$B in its ground state.
The shaded region contains events with residual excited bound or 
continuum states.  The dashed line contains events in which the $\Delta$ was excited.
Inserted is the TOF spectrum for protons detected in 
BigBite in coincidence with the $^{12}$C(e,e$^{\prime}$p) reaction. 
The random background is shown as a dashed line.
}
\end{figure}

For the highest $p_{miss}$ setting,
Fig.~3 shows the cosine of the angle, $\gamma$, between the missing 
momentum ($\vec p_{miss}$) and the recoiling proton detected in BigBite
($\vec p_{rec}$).
We also show in Fig.~3 the angular correlation for the random background as defined
by a time window off the coincidence peak.  The back-to-back peak of the real triple
coincidence events is demonstrated clearly.  The curve is a result of a simulation of
the scattering off a moving pair having a center-of-mass (c.m.) momentum width of 0.136~GeV/c as discussed below.
Similar back-to-back correlations were observed for 
the other kinematic settings.
  
\begin{figure}[hbt]
\centerline{\epsfig{width=\linewidth,file=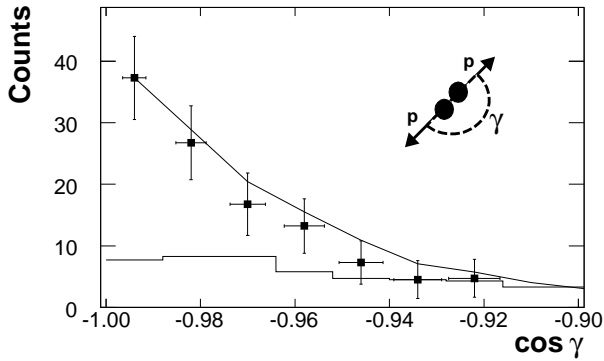}}
\caption{ The distribution of the cosine of the opening angle between the 
$\vec p_{miss}$ and $\vec p_{rec}$ for the 
$p_{miss}=0.55$~GeV/c  kinematics.  The histogram shows the distribution of random events.
The curve is a simulation of the scattering off a moving pair with a width of 0.136~GeV/c
for the pair c.m. momentum.}
\end{figure}

In the plane-wave impulse approximation (PWIA) the c.m. momentum of a pp-SRC pair is given by:  
\begin{equation}
\vec p_{c.m.} \equiv
\vec p_{miss} + \vec p_{rec}.
\label{eq:cm}
\end{equation}
For the triple-coincident events, 
we reconstructed the two components of $\vec p_{c.m.}$ 
in the direction towards BigBite 
and vertical to the scattering plane.  In these directions
the acceptance was large enough
to be sensitive to the magnitude of the c.m. motion.

To avoid distortions 
due to the finite acceptance of BigBite, we compared the measured distributions of these components to 
simulated distributions that were produced using MCEEP \cite {MCEEP}. 
The finite angular and momentum acceptances of BigBite were modeled in the 
simulation by applying the same cuts on the recoiling protons as were applied
to the data. 
The simulations assume that an electron scatters off a moving pp pair with a c.m. 
momentum relative to the A-2 spectator system described by a Gaussian distribution,
as in \cite{CiofidegliAtti:1995qe}.  We assumed an isotropic 3-dimensional motion 
of the pair and varied the width of the 
Gaussian motion equally in each direction until the best agreement with the data was obtained. 
The six measured distributions (two
components in each of the three kinematic settings) yield, within uncertainties, 
the same width with a weighted average of $0.136\pm 0.020$
GeV/c. This width is consistent with the width determined from the
(p,ppn) experiment at BNL \cite{Tang:2002ww}, 
which was $0.143\pm 0.017$ GeV/c. It is also in agreement with the
theoretical prediction of 0.139 GeV/c in reference \cite{CiofidegliAtti:1995qe}.

The measured ratio of $^{12}$C(e,e$^{\prime}$pp) to $^{12}$C(e,e$^{\prime}$p) events is given by 
the ratio of events in the background-subtracted TOF peak (insert in Fig.~2) 
to those in the shaded area in the $E_{miss}$ spectrum of Fig.~2.  
This ratio, as a function of $p_{miss}$ in the $^{12}$C(e,e$^{\prime}$p) reaction, 
is shown as the full squares in the upper panel of Fig.~4.  
The uncertainties are dominated by statistical errors; the
uncertainty in separating out events from $\Delta$-production is small. 
 
\begin{figure}[ht]
\centerline{\epsfig{width=\linewidth,file=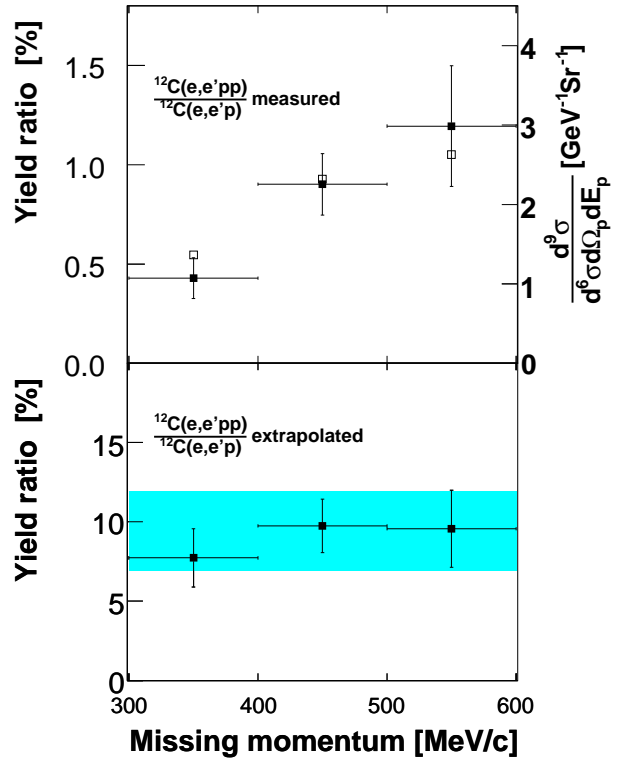}}
\caption{ The measured and extrapolated ratios 
of yields for the $^{12}$C(e,e$^{\prime}$pp) and the $^{12}$C(e,e$^{\prime}$p) reactions.  The full squares  are 
the yield ratios and the open squares are the corresponding ratios of the differential cross 
sections for the $^{12}$C(e,e$^{\prime}$pp) reaction to the $^{12}$C(e,e$^{\prime}$p) reaction. 
A simulation was used to account for the finite acceptance of BigBite and make the extrapolation 
to the total number of recoiling proton pairs shown in lower figure.  
The gray area represents a band of $\pm 2 \sigma$ uncertainty in the width of the c.m. momentum of the pair.}

\end{figure}

The measured ratio can be translated to the ratio of the nine-fold 
differential cross section for the $^{12}$C(e,e$^{\prime}$pp) reaction to the six-fold 
differential cross section for the $^{12}$C(e,e$^{\prime}$p) reaction. This ratio is  
presented as the open squares in Fig. 4.  For simplicity, the error bars
on the differential cross sections ratios are not shown because they are
very similar to those of the yield ratios.

The measured ratios in the upper panel of Fig.~4 are limited by the finite 
acceptance of BigBite.  We used the simulation described 
above to account for this finite acceptance; the resulting extrapolated ratios are shown in the lower panel of  Fig.~4.
The simulation used 
a Gaussian distribution (of width 0.136 GeV/c as determined above) for the c.m. momentum of the pp pairs.
The shaded band in the figure corresponds 
to using a width $\pm 0.040$ GeV/c (two standard deviations).  
From this result, we conclude that in the $p_{miss}$ range 
between 0.30 and 0.60~GeV/c, 
$(9.5\pm 2)\%$ of the $^{12}$C(e,e$^{\prime}$p) events have a second proton that is ejected roughly 
back-to-back to the first one, with very little dependence on $p_{miss}$.

While the detected protons are correlated in time, effects other than
pp-SRC, such as FSI, can cause the correlation.
In fact, FSIs can occur between protons in a pp-SRC pair as well as
with the other nucleons in the residual A-2 system.  Interactions
between nucleons in a pair conserve the 
isospin structure of the pair (i.e. pp pairs remain pp pairs).  Elastic FSIs 
between members of the SRC pair also do not change the c.m. momentum of the pair
as reconstructed from the momentum of the detected particles. 

The elastic (real) part of the 
FSI with the A-2 nucleons can alter the momenta, such as to make $\vec p_{miss}$ and/or $\vec p_{rec}$ and hence  
$\vec p_{c.m.}$  different from  Eqn.~1. 
The absorptive  (imaginary) part of the FSI can reduce 
the $^{12}$C(e,e$^{\prime}$pp)/ $^{12}$C(e,e$^{\prime}$p) ratio, while single charge 
exchange can turn pn-SRC pairs into $^{12}$C(e,e$^{\prime}$pp) events, 
thereby increasing the measured ratio. 
Our estimates of these FSI effects, based on a Glauber approximation using 
the method described in \cite{Mardor:1992sb}, indicate that the absorption and single charge exchange
compensate each other so that the net effect is small compared to the uncertainties in the  measurement. 
This conclusion is backed by the c.m. motion result which gives
widths for all the components that are narrow and internally consistent. 



In summary, we measured simultaneously the $^{12}$C(e,e$^{\prime}$p) and $^{12}$C(e,e$^{\prime}$pp) reactions 
in kinematics designed to maximize observation of SRCs while suppressing other effects 
such as FSIs, ICs, and MECs.
We identified directionally-correlated proton pairs in 
$^{12}$C using the $^{12}$C(e,e$^{\prime}$pp) reaction and determined 
the fraction of the $^{12}$C(e,e$^{\prime}$p) events at large $p_{miss}$ from pp-SRCs to be $(9.5\pm 2)\%$.
In the PWIA, the c.m. momentum distribution of the $pp$-SRC pair was determined 
to have a Gaussian shape with a width of $0.136\pm 0.020$ GeV/c.


We would like to acknowledge the contribution of the Hall A collaboration 
and technical staff.  Useful discussions with
J. Alster, C. Ciofi degli Atti, the late K. Egiyan,  
A. Gal, L. Frankfurt, J. Ryckebusch, M. Strikman, and
M. Sargsian, are gratefully acknowledged. 
This work was supported by the Israel Science Foundation, the US-Israeli 
Bi-national Scientific Foundation, the UK Engineering and Physical Sciences Research Council,
the U.S. National Science Foundation, the U.S. Department of Energy grants DE-AC02-06CH11357,
DE-FG02-94ER40818, and U.S. DOE Contract No. DE-AC05-84150, Modification No. M175, 
under which the Southeastern Universities Research 
Association, Inc. operates the Thomas Jefferson National Accelerator Facility.


\end{document}